\documentclass[twocolumn,twoside,slac_two]{revtex4}
\usepackage{graphicx}
\usepackage{fancyhdr}
\begin{document}

\title{Locating the $\gamma$--ray Flaring Emission of Blazar AO~0235+164 in the Jet at Parsec Scales Through Multi Spectral Range Monitoring}

\author{I.~Agudo$^{1,2}$, A.~P.~Marscher$^{2}$, S.~G.~Jorstad$^{2,3}$, V.~M.~Larionov$^{3,4}$, J.~L.~G\'omez$^{1}$, A.~L\"{a}hteenm\"{a}ki$^{5}$, P.~S.~Smith$^{6}$, K.~Nilsson$^{7}$,  A.~C.~S.~Readhead$^{8}$, M.~F.~Aller$^{9}$, J.~Heidt$^{10}$, M. Gurwell$^{11}$, C. Thum$^{12}$, A~ÊE.~Wehrle$^{13}$, O.~M.~Kurtanidze$^{14}$}
\affiliation{{$^{1}$Instituto de Astrof\'{i}sica de Andaluc\'{i}a, CSIC, Granada, Spain}\\
{$^{2}$Institute for Astrophysical Research, Boston University,Boston, USA}\\
{$^{3}$Astronomical Institute, St. Petersburg State University, St. Petersburg, Russia}\\
{$^{4}$Isaac Newton Institute of Chile, St. Petersburg Branch, St. Petersburg, Russia}\\
{$^{5}$Aalto University Mets\"{a}hovi Radio Observatory, Kylm\"{a}l\"{a}, Finland}\\
{$^{6}$Steward Observatory, University of Arizona, Tucson, USA}\\
{$^{7}$Finnish Centre for Astronomy with ESO, University of Turku, Piikki\"o, Finland}\\
{$^{8}$Cahill Center for Astronomy and Astrophysics, California Institute of Technology, Pasadena, USA}\\
{$^{9}$Department of Astronomy, University of Michigan, Ann Arbor, USA}\\
{$^{10}$ZAH, Landessternwarte Heidelberg, K\"{o}nigstuhl, Heidelberg, Germany}\\
{$^{11}$Harvard--Smithsonian Center for Astrophysics, Cambridge, USA}\\
{$^{12}$Instituto de Radio Astronom\'{i}a Milim\'{e}trica, Granada, Spain}\\
{$^{12}$Space Science Institute, Boulder, USA}\\
{$^{12}$Abastumani Observatory, Abastumani, Georgia}}

\begin{abstract}
We present observations of a major outburst at centimeter, millimeter, optical, X-ray, and $\gamma$--ray wavelengths of the BL~Lacertae object AO~0235+164 in 2008. Here we reproduce the results of the analysis of these observations, already presented in \citet{Agudo:2011p15946}. We analyze the timing of multi-waveband variations in the flux and linear polarization, as well as changes in Very Long Baseline Array (VLBA) images at 7\,mm with $\sim0.15$ milliarcsecond resolution. The association of the events at different wavebands is confirmed at high statistical significance by probability arguments and Monte-Carlo simulations. A series of sharp peaks in optical linear polarization, as well as a pronounced maximum in the 7\,mm polarization of a superluminal jet knot, indicate rapid fluctuations in the degree of ordering of the magnetic field. These results lead us to conclude that the outburst occurred in the jet both in the quasi-stationary core and in the superluminal knot, both at $>12$ parsecs downstream of the supermassive black hole. We interpret the outburst as a consequence of the propagation of a disturbance, elongated along the line of sight by light-travel time delays, that passes through a standing recollimation shock in the core and propagates down the jet to create the superluminal knot. The multi-wavelength light curves vary together on long time-scales (months/years), but the correspondence is poorer on shorter time-scales. This, as well as the variability of the polarization and the dual location of the outburst, agrees with the expectations of a multi-zone emission model in which turbulence plays a major role in modulating the synchrotron and inverse Compton fluxes.
\end{abstract}

\maketitle

\thispagestyle{fancy}

\section{Introduction}
\label{intr}
Our understanding of the processes leading to the generation of $\gamma$-ray emission from blazars, the most extreme active galactic nuclei, depends on where those $\gamma$-rays originate. 
This is currently the subject of considerable debate \citep[e.g.,][]{Marscher:2010p14415,Tavecchio:2010p14858,Agudo:2011p14707}.
Two main locations of the site of $\gamma$-ray emission in blazars have been proposed.
The first is close ($\lesssim0.1{\rm{-}}1$\,pc) to the supermassive black hole (BH).
The second one, a region much farther ($>>1$\,pc) from the BH, is beyond the ``core'' where the jet starts to be visible at millimeter wavelengths withVLBI.
Supporting this second scenario, \citet{Agudo:2011p14707} unambiguously locate the region of $\gamma$-ray flares $>14$\,pc from the BH in the jet of {OJ287} through correlation of millimeter-wave with $\gamma$-ray light curves and direct ultrahigh-resolution 7\,mm imaging with the VLBA.
Similar results are obtained by \citet{Marscher:2010p11374} and \citet{Jorstad:2010p11830} for {PKS 1510$-$089} and {3C~454.3}, respectively.

In \citet{Agudo:2011p15946} we investigate the location and properties of a radio to $\gamma$-ray outburst in the BL~Lacertae object {AO~0235+164} ({0235+164} hereafter, $z=0.94$), which results are reproduced here.

\begin{figure*}
   \centering
   \includegraphics[clip,width=13cm]{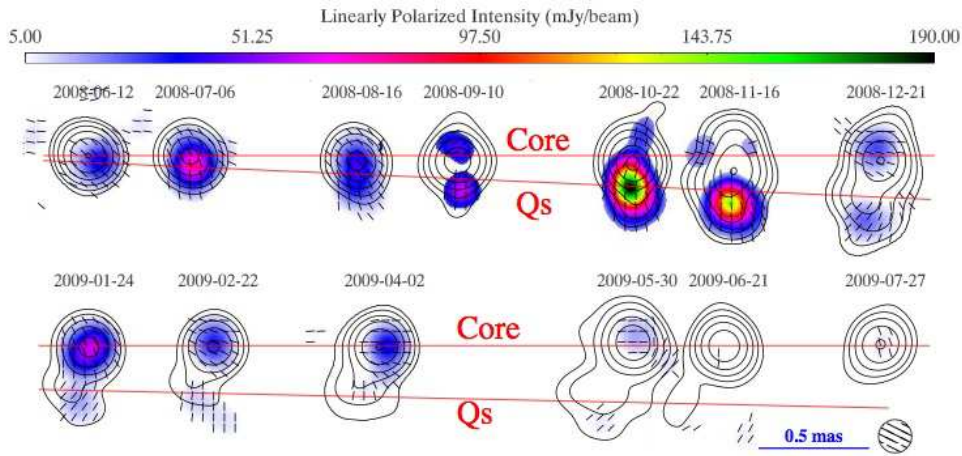}
   \caption{Sequence of 7\,mm VLBA images of 0235+164 convolved with a $\rm{FWHM}=0.15$\,mas circular Gaussian beam. Contour levels represent total intensity (levels in factors of 2 from $0.4$ to  $51.2$\,\% plus $90.0$\,\% of peak$=4.93$\,Jy/beam), color scale indicates polarized intensity, and superimposed sticks show the orientation of $\chi$. Reproduced from \citet{Agudo:2011p15946}.}
   \label{maps}
\end{figure*}

\section{Observations}
\label{obs}
Our photo-polarimetric monitoring observations of {0235+164} (Figs.~\ref{maps}-\ref{pol}) include (1) 7\,mm images with the VLBA from the Boston University monthly blazar-monitoring program, (2) 3\,mm observations with the IRAM 30m Telescope, and (3) optical measurements from the following telescopes: Calar Alto (2.2m Telescope, observations under the MAPCAT program), Steward Observatory (2.3 and 1.54m Telescopes), Lowell Observatory (1.83m Perkins Telescope), San Pedro M\'{a}rtir Observatory (0.84m Telescope), Crimean Astrophysical Observatory (0.7m Telescope), and St. Petersburg State University (0.4m Telescope). 
Our total flux light curves (Fig.~\ref{tflux}) include data from the \emph{Fermi}-LAT $\gamma$-ray (0.1--200\,GeV) and \emph{Swift}-XRT X-ray (2.4--10\,keV) observatories, available from the archives of these missions, and RXTE at 2.4--10\,keV.
Optical $R$-band fluxes come from the Tuorla Blazar Monitoring Program, the Yale University SMARTS program, and Maria Mitchell and Abastumani Observatories.
Longer wavelength light-curves were acquired from the Submillimeter Array (SMA) at 850\,$\mu$m and 1\,mm, the IRAM 30m Telescope at 1\,mm, the Mets\"{a}hovi 14m Telescope at 8\,mm, and both the Owens Valley Radio Observatory (OVRO) 40m Telescope \emph{Fermi} Blazar Monitoring Program and University of Michigan Radio Astronomy Observatory (UMRAO) 26m Telescope at 2\,cm.

We followed data reduction procedures described in previous studies: VLBA: \citet{Jorstad:2005p264}; optical polarimetric data: \citet{Jorstad:2010p11830}; IRAM data: \citet{Agudo:2006p203, Agudo:2010p12104}; SMA: \citet{Gurwell:2007p12057}; Mets\"{a}hovi: \citet{1998A&AS..132..305T}; OVRO: \citet{Richards:2010p14140}; UMRAO: \citet{Aller:1985p6715}; \emph{Swift}: \citet{Jorstad:2010p11830}; \emph{RXTE:} \citet{Marscher:2010p11374}; and \emph{Fermi}-LAT: \citet{Marscher:2010p11374, Agudo:2011p14707}. 
See \citet{Agudo:2011p15946} for further details about the X-ray and $\gamma$-ray data reduction.

\section{Major Millimeter Flare in 2008 Related to a New Superluminal Knot}
\label{ejection}
We model the brightness distribution of the source at 7\,mm with a small number of circular Gaussian components  (Fig.~\ref{maps}). 
Our model fits include a bright superluminal feature (Qs, the brightest jet feature ever detected in this object) propagating from the core at $<\beta_{\rm{app}}>=(12.6\pm1.2)\,c$.

\begin{figure*}
   \centering
   \includegraphics[clip,width=13cm]{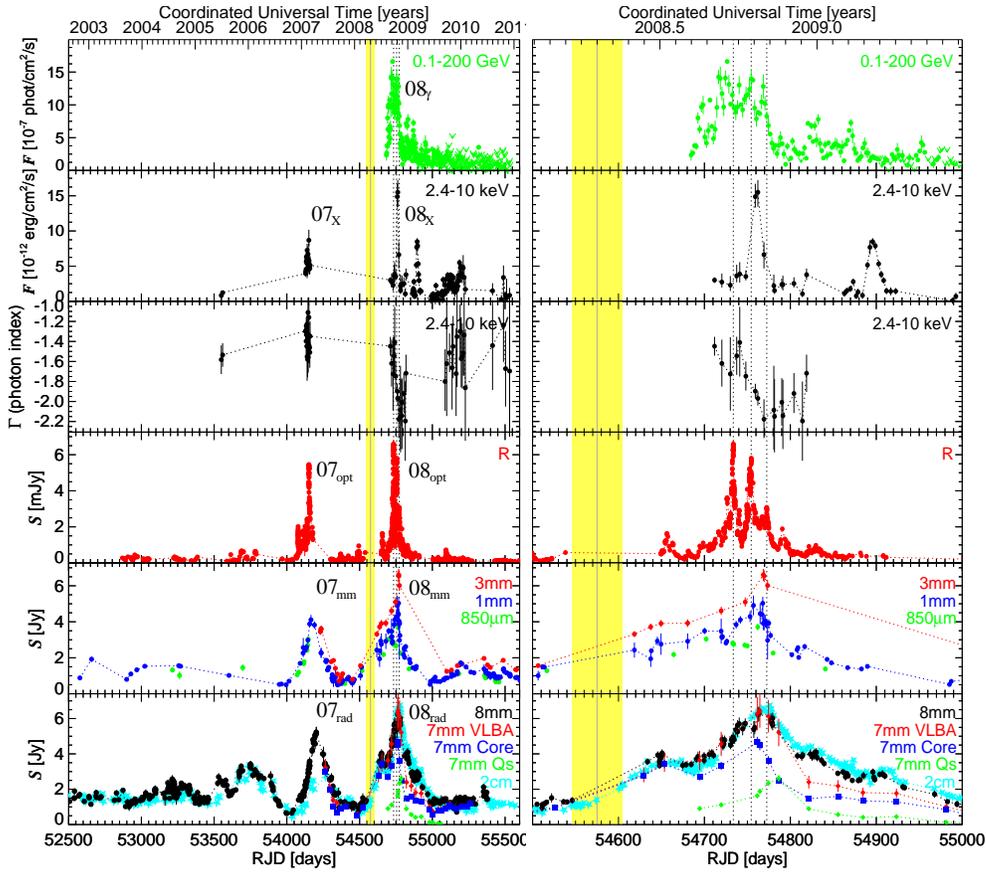}
   \caption{\emph{Left:} Light curves of 0235+164 from $\gamma$-ray to millimeter wavelengths. 
   X-ray photon index evolution from the \emph{Swift}-XRT data is also plotted. 
   Vertical dotted lines mark the three most prominent ${\rm{08}}_{\rm{opt}}$ optical peaks. The yellow area represents the time of ejection of feature Qs within its uncertainty. RJD = Julian Date~$-$~2400000.0. \emph{Right:} Same as left panel for RJD$\in[54500,55000]$. Reproduced from \citet{Agudo:2011p15946}.}
  \label{tflux}
\end{figure*}

The ejection of Qs was coincident with the core near the start of an extreme mm outburst (${\rm{08}}_{\rm{mm}}$ in Fig.~\ref{tflux})). 
Figure~\ref{tflux} shows that radio and mm outbursts in 2008 (${\rm{08}}_{\rm{rad}}$ and ${\rm{08}}_{\rm{mm}}$) contain contributions from both the core and Qs. 
Their contemporaneous co-evolution suggests that the disturbance responsible for the ejection of Qs extended from the location of the core to Qs in the frame of the observer, which could have resulted from light-travel delays \citep[e.g.,][]{1997ApJ...482L..33G,Agudo:2001p460,Mimica:2009p6237}.
The rarity of the ${\rm{08}}_{\rm{rad}}$, ${\rm{08}}_{\rm{mm}}$, and Qs events strongly implies that they are physically related.

The jet half--opening--angle of {0235+164} \citep[$\alpha_{\rm{int}}/2\lesssim1^\circ\kern-.35em .25$, see ][]{Agudo:2011p15946} and the average FWHM of the core measured from our 31 VLBA observing epochs in [2007,2010] ($\langle{\rm{FWHM}_{\rm{core}}}\rangle=(0.054\pm0.018)$\,mas), constrain the 7\,mm core to be at $d_{\rm{core}}=1.8 \langle{\rm{FWHM}_{\rm{core}}\rangle/\tan\alpha_{\rm{int}}}\gtrsim12$\,pc from the vertex of the jet cone.

\section{Contemporaneous Flares from $\gamma$-ray to Radio Wavelengths}
\label{corr}
\begin{figure*}
   \centering
   \includegraphics[clip,width=12cm]{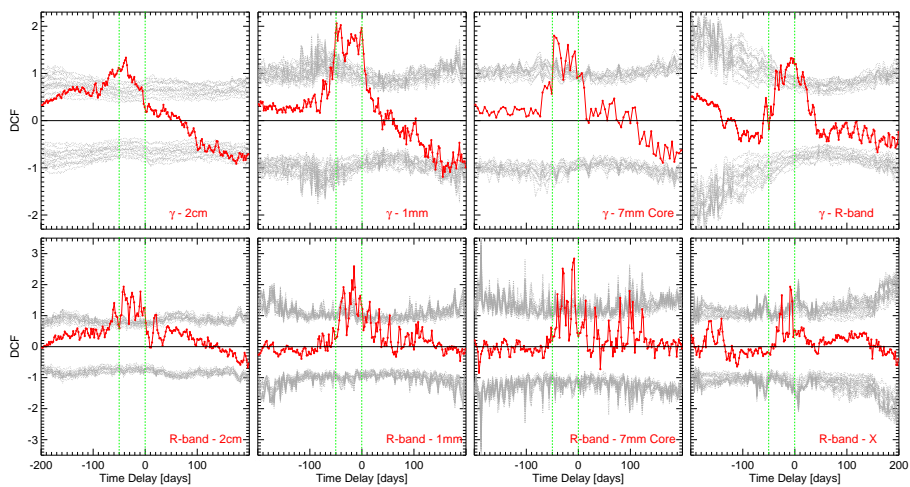}
   \caption{Grid of DCF of labeled light-curve pairs during the maximum time period ${\rm{RJD}}=[52200,55600]$. \emph{Top} row of panels show DFC with $\gamma$-ray light-curve, whereas \emph{bottom} row show DCF with $R$-band light-curve. Grey dotted curves at positive (negative) DCF values symbolize 99.7\,\% confidence limits for correlation against the null hypothesis of stochastic variability. Green dashed lines at $0$ and $-50$ day time--lags are drawn for reference. Reproduced from \citet{Agudo:2011p15946}.}
   \label{dcf1}
\end{figure*}

\begin{figure*}
   \centering
   \includegraphics[clip,width=12cm]{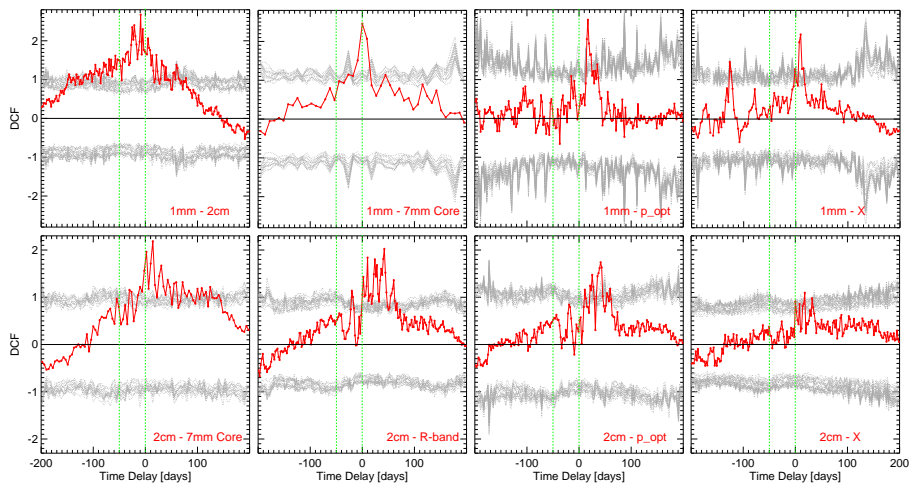}
   \caption{Same as Fig.~\ref{dcf1} but for the 1\,mm light-curve (\emph{top}) and the 2\,cm light-curve (\emph{bottom}). Reproduced from \citet{Agudo:2011p15946}.}
   \label{dcf2}
\end{figure*}

Figure~\ref{tflux} reveals that the ${\rm{08}}_{\rm{rad}}$ and ${\rm{08}}_{\rm{mm}}$ flares were accompanied by sharp optical, X--ray, and $\gamma$-ray counterparts (${\rm{08}}_{\rm{opt}}$, ${\rm{08}}_{\rm{X}}$, and  ${\rm{08}}_{\gamma}$ flares, respectively). 
Our formal light-curve correlation analysis (Fig.~\ref{dcf1}) --performed following \citet{Agudo:2011p14707}-- confirms the association of $\gamma$-ray variability with that at 2\,cm, 8\,mm, 1\,mm, and optical wavelengths at $>99.7$\,\% confidence.
The flux evolution of the VLBI core is also correlated with the $\gamma$-ray light curve at $>99.7$\,\% confidence.
Moreover, the evolution of the degree of optical linear polarization ($p_{\rm{opt}}$) and X-ray light curve are also correlated with the optical $R$-band, 1\,mm, and 2\,cm light curves at $>99.7$\,\% confidence (Fig.~\ref{dcf2}), further indicating that the extreme flaring activity revealed by our light curves is physically related at all wavebands from radio to $\gamma$-rays.

There is, however, no common pattern to the discrete correlation function (DCF) at all spectral ranges. 
This implies that, although there is correlation on long time-scales (years), on short time-scales ($\lesssim2$~months) the variability pattern does not correspond as closely.
This is the result of the intrinsic variability pattern rather than the irregular time sampling at some spectral ranges.

\begin{figure*}
   \centering
   \includegraphics[clip,width=13cm]{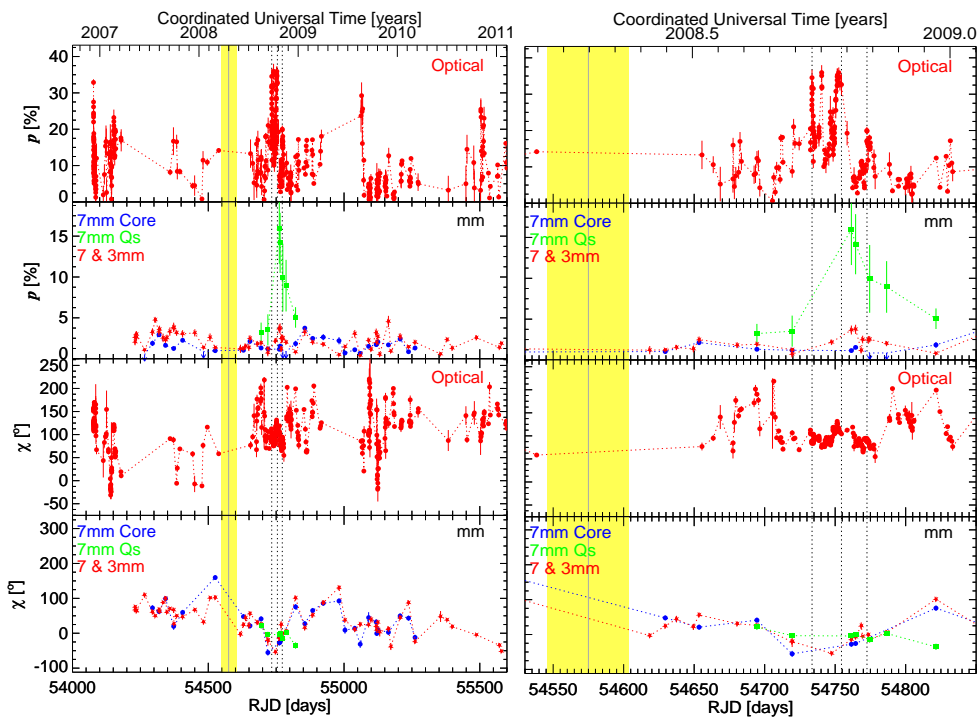}
   \caption{\emph{Left:} Long term optical and millimeter-wave linear polarization evolution of 0235+164 in the RJD=[54000,55600] range. \emph{Right:} Same as left panel for RJD$\in[54530,54850]$. Reproduced from \citet{Agudo:2011p15946}.}
   \label{pol}
\end{figure*}

\section{Correlated Variability of Linear Polarization}
\label{polvar}
Figure~\ref{pol} reveals extremely high, variable optical polarization, $p_{\rm{opt}}\gtrsim30$\,\%, during the sharp ${\rm{08}}_{\rm{opt}}$ optical peaks.
Whereas the integrated millimeter-wave degree of linear polarization ($p_{\rm{mm}}$) and that of the 7~mm core remain at moderate levels $\lesssim5$\,\%, the polarization of Qs ($p_{\rm{mm,Qs}}$) peaks at the high value of $\sim16$\,\% close to the time of the second sharp optical sub-flare.
The coincidence of this sharp maximum of $p_{\rm{mm,Qs}}$ in the brightest superluminal feature ever detected in {0235+164} with the (1) high optical flux and polarization, (2) flares across the other spectral regimes, and (3) flare in the 7-mm VLBI core, implies that the ejection and propagation of Qs in {0235+164}'s jet is physically tied to the total flux and polarization variations from radio to $\gamma$-rays.

On long time-scales (years), the linear polarization angle at both optical ($\chi_{\rm{opt}}$) and millimeter ($\chi_{\rm{mm}}$ and $\chi_{\rm{mm}}^{\rm{core}}$) wavelengths varies wildly, without a preferred orientation or systematic common trend. 
However, during flare ${\rm{08}}_{\rm{opt}}$, $\chi_{\rm{opt}}$ maintains a stable orientation at $(100\pm20)^{\circ}$, whereas $\chi_{\rm{mm}}^{\rm{Qs}}$ is roughly perpendicular to this ($\sim0^{\circ}$), as expected for a plane-perpendicular shock wave propagating to the south towards Qs.
Owing to the large peak value of Qs, $p_{\rm{mm,Qs}}^{\rm{max}}\sim16$\,\%, one cannot explain the orthogonal optical-millimeter polarizations by opacity effects.
Instead, we propose that the optical polarization mainly arises in a conical shock associated with the 7-mm core, while the millimeter-wave polarization results from a propagating shock front associated with Qs. 
We surmise that the moving shock also emits polarized optical radiation, since the optical polarization drops precipitously when the orthogonal polarization of Qs peaks (Fig.~\ref{pol}-\emph{right}).


\section{Low Probability of Chance Coincidences}
\label{prob}
The relationship among the $\gamma$-ray, optical, and radio-millimeter flares is supported by probability arguments.
If the flares occur randomly, the probability of observing, at any time, a $\gamma$-ray outburst like the one reported here (i.e., with flux $\gtrsim10^{-6}\,{\rm{phot}}\,{\rm{cm}}^{-2}\,{\rm{s}}^{-1}$ and duration $\sim70$ days) is $p_{\gamma}=0.08$.
For optical and radio--millimeter wavelengths, this probability is $p_{\rm{opt}}=0.04$ and $p_{\rm{mm}}=0.15$, respectively.
Thus, if the flares at different wavelengths were random and independent of each other, the probability of observing a $\gamma$-ray, optical, and radio-millimeter flare at any given time is $p_{\gamma,{\rm{opt}},{\rm{mm}}}=5\times10^{-4}$. 
This counters the null hypothesis of random coincidence at 99.95\,\% confidence.

\section{Observational Evidences}
The coincidence of the ejection and propagation of Qs --by far the brightest non-core feature ever reported in {0235+164}-- with the prominent $\gamma$-ray to radio outbursts and the extremely high values of $p_{\rm{opt}}$ and $p_{\rm{mm,Qs}}$ provides convincing evidence that all these events are physically connected. 
This is supported by probability arguments and by our formal DCF analysis, which unambiguously confirms the relation of the $\gamma$-ray outburst in late 2008 with those in the optical, millimeter-wave (including the 7-mm VLBI core) and radio regimes.
We locate the millimeter core at $d_{\rm{core}}\gtrsim12$\,pc from the vertex of the jet.

\section{Conclusions}
We identify the 7-mm core as the first re-collimation shock near the end of the jet's acceleration and collimation zone \citep[ACZ][]{2007AJ.134.799J,Marscher:2008p15675,Marscher:2010p11374}.
Superluminal feature Qs is consistent with a moving shock oriented transverse to the jet axis, given the extremely high $p_{\rm{mm,Qs}}$, with $\chi_{\rm{mm}}^{\rm{Qs}}$ parallel to the direction of propagation of Qs.
The flux evolution of the core appears closely tied to that of Qs, and its light curve is correlated at high confidence with those at $\gamma$-ray, optical, and millimeter wavelengths.
This suggests that Qs is the head of an extended disturbance, perhaps containing a front-back structure stretched by light-travel delays in the observer's frame \citep[see, e.g.,][]{Aloy:2003p350}.
 
\begin{figure*}
   \centering
   \includegraphics[clip,width=16cm]{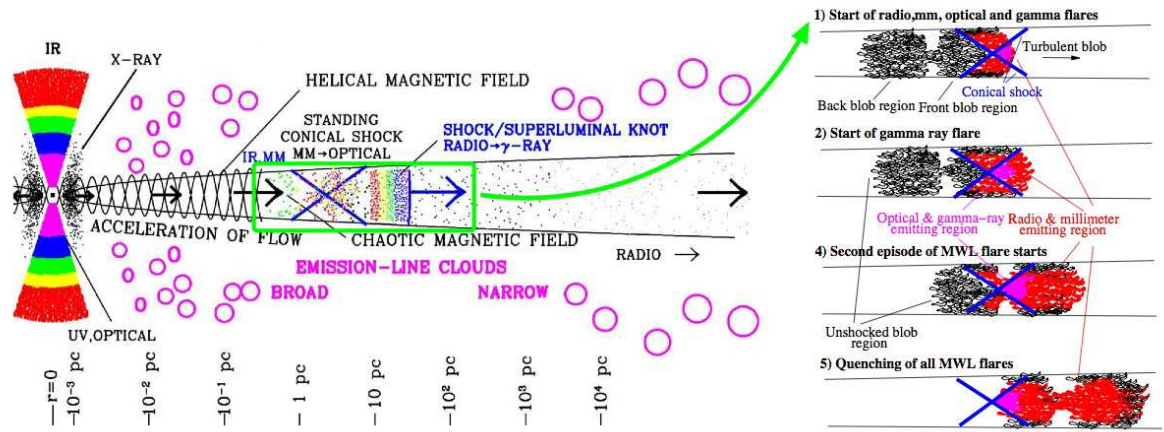}
   \caption{Scheme of proposed model for the multi-wavelength flaring behavior of AO~0235+164. Radio-loud AGN sketch adapted from \citet{Marscher:2006p362}.}
   \label{sketch}
\end{figure*}

Under this scenario, the radio/millimeter-wave and optical (and perhaps X-ray) synchrotron flares start when the front region crosses the conical shock at the core, where the jet is at least partially optically thin (Fig.~\ref{sketch}).
This interaction accelerates electrons, and produces inverse Compton $\gamma$-ray emission from the up-scattering of IR-optical photons. 
When the back region of the moving perturbation encounters the core, their interaction again produces efficient particle acceleration, which is seen as a sudden optical and radio/millimeter synchrotron emission enhancement.
The subsequent optical variability is produced by the passage of the remaining shocked turbulent plasma in the back structure through the core.
During the different optical sub--flares, the integrated radio/millimeter synchrotron flux keeps rising.
This radio/millimeter outburst is more prolonged owing to the longer synchrotron cooling-time of electrons radiating at these wavelengths and the lower speed of the back structure.
Indeed, Qs does not reach its maximum radio/millimeter-wave flux until traveling a projected distance of $\sim0.13$ mas from the core.
When the entire front-back structure passes across the core, the synchrotron emission declines rapidly at optical (and, if relevant, X-ray) frequencies, as does the $\gamma$-ray emission.
The decay of the radio/millimeter-wave emission is more gradual (see above).

\bigskip 
\begin{acknowledgments}
We acknowledge the anonymous referee for constructive comments.
This research was funded by NASA grants NNX08AJ64G, NNX08AU02G, NNX08AV61G, and NNX08AV65G, NSF grant AST-0907893, and NRAO award GSSP07-0009 (Boston University); RFBR grant 09-02-00092 (St.~Petersburg State University); MICIIN grant AYA2010-14844, and CEIC (Andaluc\'{i}a) grant P09-FQM-4784 (IAA-CSIC); the Academy of Finland (Mets\"{a}hovi); NASA grants NNX08AW56S and NNX09AU10G (Steward Observatory); and GNSF grant ST08/4-404 (Abastunami Observatory).
The VLBA is an instrument of the NRAO, a facility of the NSF under cooperative agreement by AUI. 
The PRISM camera was developed by Janes et~al., and funded by NSF, Boston University, and Lowell Observatory. 
Calar Alto Observatory is operated by MPIA and IAA-CSIC. 
The IRAM 30m Telescope is supported by INSU/CNRS, MPG, and IGN.
The SMA is a joint project between the SAO and the Academia Sinica. 
\end{acknowledgments}


\end{document}